\documentclass[runningheads]{llncs}
\usepackage{graphicx}
\usepackage{booktabs}
\usepackage{multirow}
\usepackage{todonotes}
\usepackage{hyperref}
\usepackage{adjustbox}
\usepackage{makecell}

\makeatletter
\newcommand{\printfnsymbol}[1]{
  \textsuperscript{\@fnsymbol{#1}}}
\usepackage{cleveref}
\begin{document}
\title{Extending nnU-Net is all you need}
%
%
\author{Fabian Isensee\thanks{equal contribution} \inst{1,2}\and
Constantin Ulrich\printfnsymbol{1}\inst{1} \and
Tassilo Wald\printfnsymbol{1}\inst{1,2} \and Klaus Maier-Hein \inst{1,2}}
\authorrunning{Isensee F., Ulrich C., Wald T. \& Maier-Hein K.H.}
%
\institute{Division of Medical Image Computing, German Cancer Research Center (DKFZ)\
\and Helmholtz Imaging}
\maketitle              
\begin{abstract}
Semantic segmentation is one of the most popular research areas in medical image computing. Perhaps surprisingly, despite its conceptualization dating back to 2018, nnU-Net continues to provide competitive out-of-the-box solutions for a broad variety of segmentation problems and is regularly used as a development framework for challenge-winning algorithms. Here we use nnU-Net to participate in the AMOS2022 challenge, which comes with a unique set of tasks: not only is the dataset one of the largest ever created and boasts 15 target structures, but the competition also requires submitted solutions to handle both MRI and CT scans. 
Through careful modification of nnU-net's hyperparameters, the addition of residual connections in the encoder and the design of a custom postprocessing strategy, we were able to substantially improve upon the nnU-Net baseline. Our final ensemble achieves Dice scores of 90.13 for Task 1 (CT) and 89.06 for Task 2 (CT+MRI) in a 5-fold cross-validation on the provided training cases.


\end{abstract}
\section{Introduction}
    Automated delineation of all anatomical structures and pathologies in medical images is a long-standing goal in medical image computing. Due to the need of expert annotators and the time-intensive nature of 3D annotations, datasets have so far required careful balancing between the number of target structures and the number of training cases. 
    Consequently, existing methods are either trained on many images and can robustly segment few structures or are trained on few images can segment many structures with reduced robustness. Thus, whenever a holistic perspective on a patients anatomy is required, multiple expert models must be pooled together from multiple sources. Not only does this increase the inference time, but it also creates new issues such as potentially conflicting predictions. Finally and perhaps most importantly, with each expert model being trained independently, label synergies cannot be exploited, thus potentially decreasing the label efficiency of the models as well as their robustness. \\
    In this context, the Abdominal Multi Organ Segmentation 2022 (AMOS2022) challenge \cite{ji2022amos}, is set out to catalyzes the development of holistic segmentation methods. It comes with an unprecedented number of training images and annotated target structures: 15 organs of interest were labeled in 500 CT and 100 MRI scans, distributed into 200+40 (CT + MRI) training, 100+20 validation and 200+40 test images. The challenge poses two tasks: Task 1 is a classic multi-organ segmentation problem on just the CT images whereas Task 2 includes the MRI images and expects submitted methods to handle both modalities.
    
    Within the context of medical image segmentation, nnU-Net \cite{isensee_nnu-net_2021} has stood the test of time. It consistently delivers state-of-the-art results on new segmentation datasets as they are released, despite its fully automated out-of-the-box nature. Moreover, nnU-Net was successfully used as a basis for task-specific method optimization, enabling not just us \cite{isensee2020nnu,full2020studying} but also many other teams \cite{ma2021cutting} to win highly contested challenges. Thus, it seems only natural to use nnU-Net for our participation in the AMOS2022 challenge as well.

\section{Method}
    nnU-Net is a framework that automatically configures and trains U-Net \cite{ronneberger_u-net_2015} based segmentation pipelines. Through rigorous analysis of the target dataset, nnU-Net makes automated adaptations to the patch size, batch size, preprocessing, network topology and more. For a full description of nnU-Net, we refer to \cite{isensee_nnu-net_2021}.
    
    In this section we propose several modifications to nnU-net's automatically generated pipeline to maximize segmentation performance on the AMOS2022 challenge. 
    Throughout method development we apply all modifications to both tasks with the sole difference being the intensity normalization scheme. For Task 1 we utilize nnU-Net's 'CT' scheme (data-driven clipping and normalization) and for Task 2 we use simple z-scoring of all images (nnU-Net's 'nonCT' setting).

    \subsection{Optimization of nnU-Net's segmentation pipeline to AMOS2022} 
        \label{sec:training_pipeline}
\begin{figure}[ht]
            \centering
            \includegraphics[width=.7\linewidth]{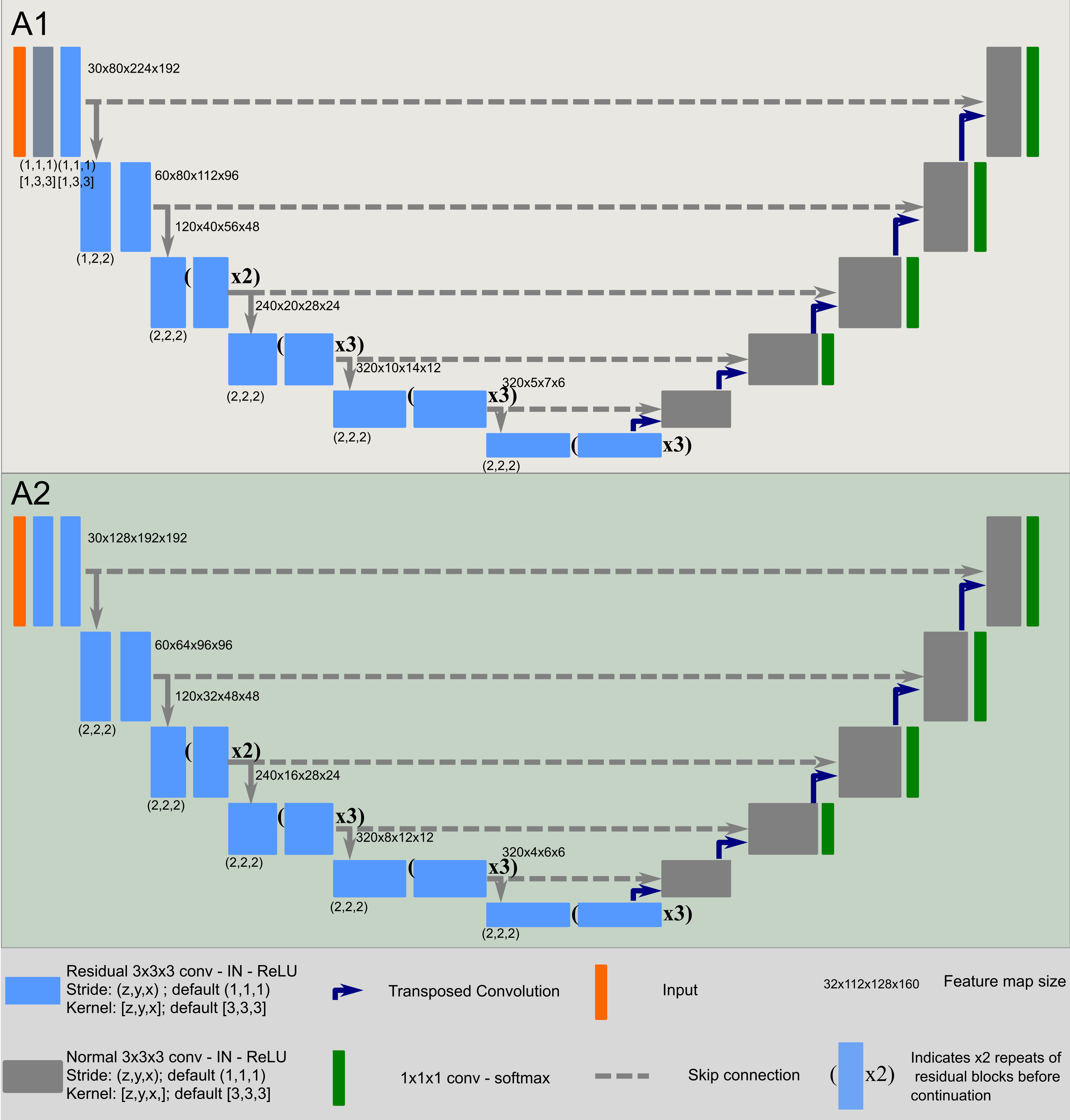}
            \caption{The two architectures used by our final configurations. Due to nnU-Net's automatic configuration of kernel sizes and strides as a function of the patch size, their feature map sizes differ. Both architectures make use of residual connections in the encoder.}
            \label{fig:amos_architectures}
        \end{figure}

        Starting from the default '3d\_fullres' configuration provided by nnU-Net we explore multiple improvements. We experimented with replacing the default encoder of the U-Net with a residual encoder (based on \cite{He2016}). We furthermore optimized the preprocessing, specifically the batch size, patch size and target spacing. Advances in GPU memory capacity and processing speed allowed for larger models and batch sizes than the standard nnU-Net. Method development is performed by running 5-fold cross-validation on the provided training cases. All models are trained from scratch using the nnU-Net framework.
        Our experimentation resulted in three well-performing candidates for Task 1 and two candidates for Task 2, all of which are summarized in \cref{tab:pipelines}.
        \begin{table}[h]
            \centering
            \caption{Final configurations used in our submission. Table highlights changes to the nnU-Net defaults.}
            \label{tab:pipelines}
            \begin{adjustbox}{width=\linewidth,center}
            \begin{tabular}{cccccccc}
                \toprule
                Task & Name & Patch Size & Spacing [mm] & Data Aug. & batch size & norm. & Arch.\\
                \midrule
                - & '3d\_fullres' & [64,160,160] & [2,0.69,0.69] & default & 2 & CT/z-score & - \\
                \midrule
                \multirow{3}{*}{1} &  Configuration 1 & [128,192,192] & [1.5,1,1] & DA5 & 5 & CT & A2\\
                & Configuration 2 & [80,224,192]  & [2,0.69,0.69]  & default & 6 & CT & A1\\
                & Configuration 3 & [128,192,192] & [1.5,1,1] & default & 5 & CT & A2\\
                \midrule
                \multirow{2}{*}{2} & Configuration 4 & [80,224,192] & [2,0.69,0.69] & default & 6 & z-score  & A1\\
                & Configuration 5 & [128,192,192] & [1.5,1,1] & default & 5 & z-score & A2\\ 
                \bottomrule
            \end{tabular}
            
            \end{adjustbox}
        \end{table}
        Fig. \ref{fig:amos_architectures} shows the segmentation architectures used by our final configurations. They share the same topology but their feature map sizes differ due to nnU-Net's automatic configuration of convolutional strides and kernel sizes as a function of the patch size.

    \subsection{Inference strategy}
        Prediction is carried out with the nnU-Net defaults (sliding window). For validation and test set prediction we use ensembling, using both the 5 models from our cross-validation as well as multiple different configurations (3 configurations for Task 1: 3 x 5 = 15 models in the ensemble). Ensembling is implemented as simple averaging of softmax outputs.

    \subsection{Postprocessing}  
        \label{sec:postprocessing}
        The postprocessing offered by nnU-net was designed to cover a wide variety of use-cases. We believe that additional performance can be gained on AMOS2022 by specifically analyzing and targeting failure cases of our method. To identify those we generated a confusion matrix on our cross-validation results (see supplementary information \cref{fig:confusion_matrix}) and performed rigorous visual inspection of our predicted segmentation maps. 
        
        \noindent \textbf{Left-right confusion.}
            Sometimes parts of the left kidney were classified as right kidney and vice-versa (same for adrenal glands). Here, removing disconnected small components is suboptimal because then parts of the kidneys would be labeled as background. 
            Thus, we pooled the final kidney predictions into a joint class and use a connected component analysis to determine connected kidney regions. For each connected region we calculate its position within the image and assign its label (left/right) accordingly. 

        \noindent \textbf{Connected component filtering.}
            Just like the default nnU-Net we explore filtering of connected components. Specifically we determine whether removing all but the largest connected component can improve the Dice score.

        \noindent \textbf{Organ size constraints.}
            Driven by the human anatomy, organs are expected to have a certain volume. This property can be exploited to filter small false positive predictions, for example in images in which the target organ is not present but false positive areas are. We use a parameter called 'rate' in combination with the minimum organ size observed in the training set. The rate is intended to be a safety margin: a rate of 0.75 indicates that components smaller than 75\% of the minimum organ size are removed. 
            
        \noindent \textbf{Component filtering with size constraints.}
            We combine the two previous techniques and remove instances below a volume threshold unless its the only connected region.
            
        \noindent \textbf{Composition of post-processing steps.}
            In order to find the optimal post-processing scheme we create compositions of the previously mentioned post-processing schemes, namely:
            \begin{enumerate}
                \item PP1: Left right confusion (only kidneys \& adrenal glands) followed by organ size constraints 
                \item PP2: Left right confusion (only kidneys \& adrenal glands) followed by component filtering with size constraints
                \item PP3: Left right confusion (only kidneys \& adrenal glands) followed by connected component filtering.
                \item PP4: Only left right confusion (only kidneys \& adrenal glands).
            \end{enumerate}
            
            We use the predictions of our ensemble on the 5-fold cross-validation (training images) to optimize the postprocessing scheme. Each organ is optimized independently and the best performing postprocessing strategy is retained. Note that we test multiple possible values for the 'rate' parameter where applicable. We refer to supplementary tables \ref{tab:postprocessing} and \ref{tab:postprocessing_task2} for the final postprocessing configurations.

\section{Results} 
    \label{sec:quantitative_results}
    The proposed modifications to the default nnU-Net pipeline substantially improved the results both on the training set cross-validation as well as the official validation set. We perform ablation studies to highlight the contributions of each component, see \cref{res_table}. 
    Overall, we were able to increase the Dice for configuration 3 (see \cref{tab:pipelines}) by about 1\% w.r.t. the nnU-Net baseline by modifying nnU-Net's automated configuration (specifically, target spacing, patch size) and using residual connections in the encoder of the U-Net. Increasing the batch size yielded further improvements.

    \begin{table}[h]
        \centering
        \caption{Ablation results.}
        \label{res_table}
        \begin{tabular}{lcccc}
                \toprule
                \multirow{2}{*}{Method} & \multicolumn{2}{c}{Task 1} & \multicolumn{2}{c}{Task 2} \\
                \cmidrule{2-5} \\
                {} & 5-fold CV & Val & 5-fold CV & Val \\ 
                {} & Dice & \makecell{mean score \\(Dice score)} & Dice & \makecell{mean score \\Dice score} \\
                \midrule
                nnU-Net default & 88.64 & \makecell{86.52 \\ (90.31)} & - & - \\ 
                \midrule
                + configuration improvements & 89.08 & - & - & - \\
                \midrule
                + residual encoder  & 89.45 & - & 88.43 & - \\ 
                \midrule
                \makecell{+ increased batch size (bs 5) \\ (corresponds to configs 3 \& 5)} & 89.57 & \makecell{87.71 \\ (91.43)} &  88.68 & \makecell{87.72 \\ (90.97)}   \\
                \bottomrule
        \end{tabular}
    \end{table}
    
    For the sake of brevity we do not show detailed results of the remaining configurations. We summarize the cross-validation performance of all our configurations and their ensembles in \cref{model_res_table}. As expected, ensembling provided a substantial gain in segmentation performance, as did our postprocessing.

    \begin{table}
        \centering
        \caption{5-fold cross-validation results of all our configurations and their ensembles.}
        \label{model_res_table}
                \begin{tabular}{lc@{\hspace{0.5cm}}lc}
                \toprule
                \multicolumn{2}{c}{Task 1} & \multicolumn{2}{c}{Task 2}\\
                Config. & Dice & Config. &Dice \\
                \midrule
                Config. 1& 89.60 & Config. 4 & 88.56\\
                \midrule
                Config. 2& 89.59 & Config. 5 & 88.69\\
                \midrule
                Config. 3& 89.57 & {} & {}\\
                \midrule
                Ensemble & 89.92 & Ensemble & 88.94\\
                + postprocessing & \textbf{90.13} & + postprocessing & \textbf{89.06}\\
                \bottomrule
                \end{tabular}
    \end{table}

\section{Discussion}
    In this paper we performed task-specific optimizations of the default nnU-Net pipeline to maximize segmentation performance on the AMOS2022 challenge. Changing crucial hyperparameters such as the patch size, batch size and target spacing for resampling yielded substantial gains relative to the default configuration, as did the addition of residual connections in the encoder of the U-Net. Our final submission consists of three configurations for Task 1 and two for Task 2. Since each configurations was trained as a 5-fold cross-validation, our ensembles consist of 15 and 10 models, respectively.
    At the time of submission, we ranked third in Task 1 and first in Task 2, although we should add that none of these submissions are reflected in this paper since we decided on short notice to deviate from our original plan and use larger ensembles instead (which according to our testing should take approx. 60h on Task 1 and 50h on Task 2 for inference on an RTX 3090). This decision was made after the validation set submission was closed and could thus only be evaluated on the training set cross-validation. Nonetheless, we are confident in the capabilities of our solution and hope for the improved performance we measured to translate into the test set.
    Naturally, the source code for training our models as well as the inference dockers will be made publicly available after the competition.
    
\section*{Acknowledgements}
Part of this work was funded by Helmholtz Imaging, a platform of the Helmholtz Incubator on Information and Data Science.

\bibliographystyle{splncs04}
\bibliography{bibfile.bib}

\newpage
\section*{Appendix}
    \begin{figure}
        \centering
        \includegraphics[width=.55\linewidth]{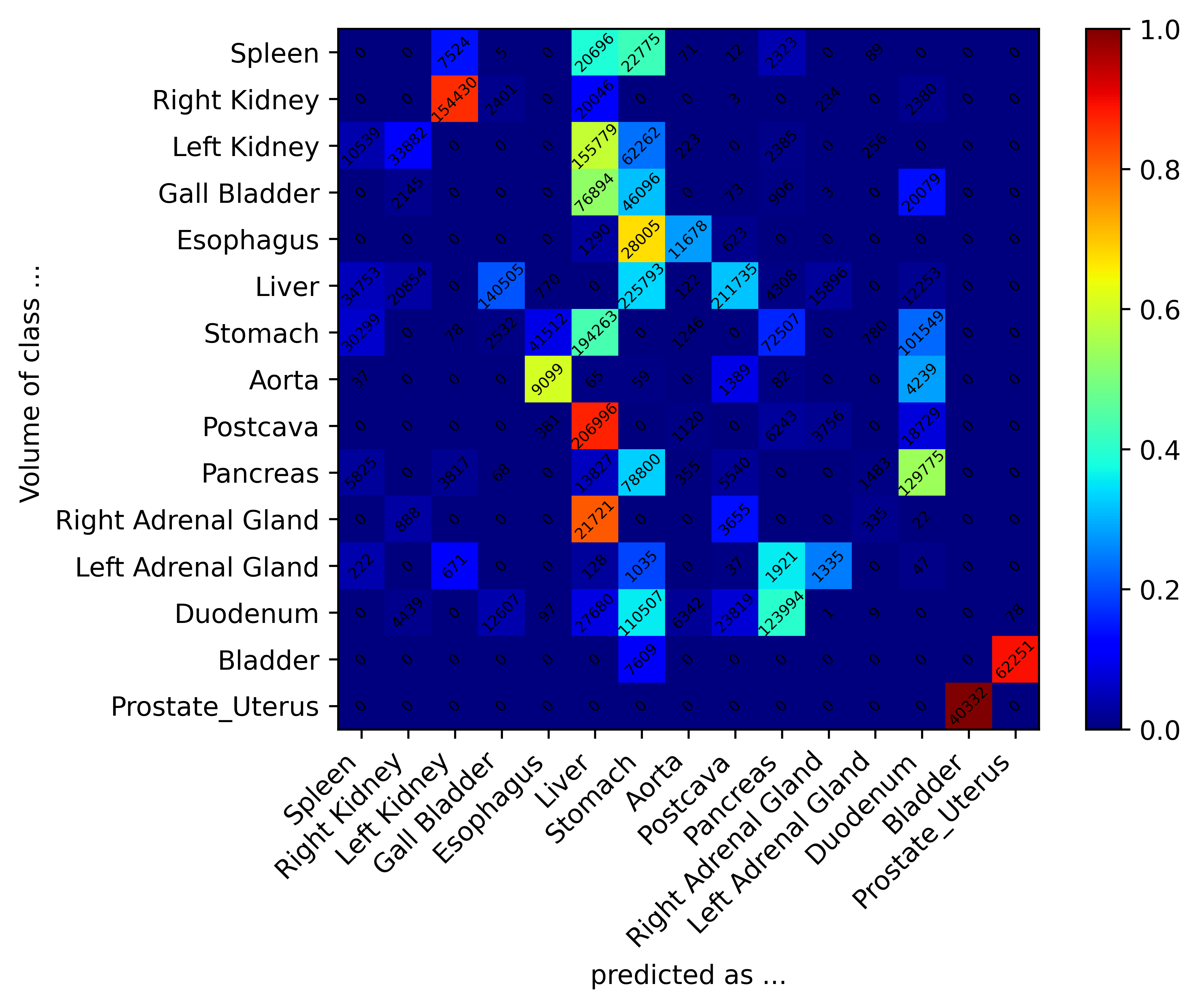}
        \caption{Confusion of organs. The color encodes the relative percentage per row while the number represents the absolute volume over the entire cross validation dataset. We used this as an indicator which error cases exist and could be solved through postprocessing. The diagonal was removed to not influence percentage calculations.}
        \label{fig:confusion_matrix}
    \end{figure}

    \begin{table}
        \centering
        \caption{Postprocessing results for the different organs. We exemplarily show the results for our final ensemble on Task 1 as calcualted on our cross validation. As previously explained, the rate indicates at which percentage of the minimum volume of the organ, a connected prediction of that organ gets removed.}
        \label{tab:postprocessing}
        \begin{tabular}{lccccc}
             \toprule
             Organ & Min. Organ Vol. [mm³] & best PP method & rate & \multicolumn{2}{c}{Dice}\\ \cmidrule{5-6}
             {} &  {} & {} & {} & before PP & after PP\\
             
             \midrule
            Spleen & 14514.5 & PP1 & 0.9 & 97.104 & \textbf{97.624} \\
            R. Kidney & 32323.3 & PP1 & 0.10 & 96.516 & \textbf{96.753} \\
            L. Kidney & 10084.4 & PP3 & - & 95.443 & \textbf{96.277} \\
            Gall Bladder & 1242.1 & PP1 & 0.25 & 86.014 & \textbf{87.360} \\
            Esophagus & 418.9 & PP1 & 0.25 & 85.480 & \textbf{85.481} \\
            Liver & 653406.2 & PP3 & - & 97.880 & \textbf{97.944} \\
            Stomach & 30.6 & PP1 & 0.9 & 91.580 & \textbf{91.579} \\
            Aorta & 26481.4 & PP1 & 0.75 & 95.901 & \textbf{96.036} \\
            Postcava & 31401.8 & PP1 & 0.10 & 91.745 & \textbf{91.755} \\
            Pancreas & 15213.3 & - & - & \textbf{86.734} & 86.734 \\
            R. Adrenal Gland & 1076.1 & - & - & \textbf{79.689} & 79.689 \\
            L. Adrenal Gland & 663.7 & PP1 & 0.1 & 80.996 & \textbf{81.010} \\
            Duodenum & 19244.0 & - & - & \textbf{83.789} & 83.789 \\
            Bladder & 14665.4 & PP1 & 0.25 & 92.285 & \textbf{92.304} \\
            Prostate/Uterus & 7937.5 & PP1 & 0.25 & 87.610 & \textbf{87.610} \\
            \midrule
            Average &  & & & 89.918 & \textbf{90.130}\\
             \bottomrule
        \end{tabular}
    \end{table}
            
    \begin{table}
        \centering
        \caption{Postprocessing results for the different organs. Equivalent for Task 2 to \cref{tab:postprocessing}.}
        \label{tab:postprocessing_task2}
        \begin{tabular}{lccccc}
             \toprule
             Organ & Min. Organ Vol. [mm³] & best PP method & rate & \multicolumn{2}{c}{Dice}\\ \cmidrule{5-6}
             {} &  {} & {} & {} & before PP & after PP\\
             
             \midrule
            Spleen & 14514.5 & PP1 & 0.10 & 96.915 & \textbf{97.325} \\
            R. Kidney & 32323.3 & - & - & \textbf{96.408} & 96.408 \\
            L. Kidney & 10084.4 & - & - & \textbf{95.606} & 95.606 \\
            Gall Bladder & 1242.1 & PP1 & 0.5 & 85.981 & \textbf{86.387} \\
            Esophagus & 418.9 & PP1 & 0.95 & 82.792 & \textbf{82.798} \\
            Liver & 653406.2 & PP3 & - & 97.865 & \textbf{97.963} \\
            Stomach & 30.6 & PP1 & 0.25 & 91.161 & \textbf{91.161} \\
            Aorta & 26481.4 & PP2 & 0.9 & 95.392 & \textbf{95.495} \\
            Postcava & 31401.8 & PP1 & 0.10 & 91.252 & \textbf{91.273} \\
            Pancreas & 15213.3 & - & - & \textbf{86.660} & 86.660 \\
            R. Adrenal Gland & 1076.1 & - & - & \textbf{77.002} & 77.002 \\
            L. Adrenal Gland & 663.7 & - & - & \textbf{78.007} & 78.007 \\
            Duodenum & 19244.0 & PP1 & 0.10 & \textbf{81.935} & 81.948 \\
            Bladder & 14665.4 & PP1 & 0.50 & 90.430 & \textbf{90.795} \\
            Prostate/Uterus & 7937.5 & PP1 & 0.25 & 86.658 & \textbf{87.121} \\
            \midrule
            Average & & & & 88.938 & \textbf{89.064} \\
             \bottomrule
        \end{tabular}
    \end{table}
\end{document}